# Modular and Landen transformation between two kinds of separable finite solutions of Sine-Gordon equation with phase N=2

**Asbrast**

In [9], we get the transformations between finite solutions of Sine-Gordon Equation (SGE) with phase=1. In this article, we will study on the next easier case, the separable N=2 solutions of Sine-Gordon equation (SGE) because we not yet understand the two-body problem in this non-linear equation in classical level. The separable case can divide the system into several independent "N=1" case by transformation. Then we can use " action and angle" principle in classical mechanics to study them. From the original symmetry, we get two kinds of N=2 separable solutions appear in [1]. we find these two kinds are related to Landen transformation.

**Nan Hong Kuo**

**C. D. Hu**

**Department of Physics, National Taiwan Univerity, Taiwan, Republic of China**

## 1. Considering original symmetry of N=2 solutions:

### 1.1 Spatial (anti)symmetry ⇔ Spectral symmetry:

We begin with the Takhatajian-Faddeev equation, which is equivalent to Sine-Gordon equation (SGE). We study their discrete eigenstates, including even function, the breathers and odd function, the kinks. In **Sec. VII** of **Ref. [4]**, the author prove spatial symmetry ⇔ spectral symmetry in the even case. In this section, we include the odd case.

**Lemma** *the Takhatajian-Faddeev eigenvalue problem:*

$$[\begin{pmatrix} 0 & -1 \\ 1 & 0 \end{pmatrix}]\frac{d}{dx} + \frac{i}{4}(v+u_x)\begin{pmatrix} 0 & 1 \\ 1 & 0 \end{pmatrix} + \frac{1}{16\sqrt{E}}\begin{pmatrix} e^{iu} & 0 \\ 0 & e^{-iu} \end{pmatrix} - \sqrt{E}]\phi = 0 \quad\quad 1$$

with $v = u_t$, and well known $\phi_{xt} = \phi_{tx} \Leftrightarrow u_t + u_x = w$; $w_t - w_x = -\sin u \Leftrightarrow u_{tt} - u_{xx} + \sin u = 0$, which is SGE. If

$$(u(-x), v(-x)) = (u(x), v(x)) \quad\quad (2)$$
$$u(x+L) = u(x) + 2\pi M, M \equiv \text{"charge" of } u(x)$$
$$v(x+L) = v(x)$$

then

$$E_j \in \sum\nolimits^{(b)} \Rightarrow \frac{1}{16^2 E_j} \in \sum\nolimits^{(b)} \quad\quad (3)$$

Consider any solution $\phi = \begin{pmatrix} \phi_1(x,E) \\ \phi_2(x,E) \end{pmatrix}$ of (1), we define the function

$\psi(x,E) = \begin{pmatrix} e^{\frac{i}{4}u(x)}\phi_1(x,E) \\ e^{-\frac{i}{4}u(x)}\phi_2(x,E) \end{pmatrix} = \begin{pmatrix} \psi_1(x,E) \\ \psi_2(x,E) \end{pmatrix}$, put into (1), then $\psi$ satisfies the eigenvalue problem:

$$-\frac{d\psi_2(x,E)}{dx} + \frac{i}{4}v \cdot \psi_2(x,E) + (\frac{1}{16\sqrt{E}}e^{\frac{i}{2}u(x)} - \sqrt{E}e^{-\frac{i}{2}u(x)})\psi_1(x,E) = 0 \qquad 4.1$$

$$\frac{d\psi_1(x,E)}{dx} + \frac{i}{4}v \cdot \psi_1(x,E) + (\frac{1}{16\sqrt{E}}e^{-\frac{i}{2}u(x)} - \sqrt{E}e^{\frac{i}{2}u(x)})\psi_2(x,E) = 0 \qquad 4.2$$

**Case 1**:

If $u$ is an even functions ( p.s. $u(x) = u(-x)$, breather):

(*pf*) : If $\begin{pmatrix}\psi_1(x,E)\\ \psi_2(x,E)\end{pmatrix}$ and $E$ are eigenfunction and value of (4); then

$\begin{pmatrix}\psi_1(-x,E)\\ \psi_2(-x,E)\end{pmatrix} = \begin{pmatrix}\widetilde{\psi}_2(x,\widetilde{E})\\ \widetilde{\psi}_1(x,\widetilde{E})\end{pmatrix}$ where $\widetilde{E} = \frac{1}{16^2 E}$, are eigenfunction and value of (4).

Put above assumation into (4), we obtain

$$-\frac{d}{dx}\widetilde{\psi}_1(-x,\widetilde{E}) + \frac{i}{4}\frac{d}{dt}u(-x) \cdot \widetilde{\psi}_1(-x,\widetilde{E}) + (\frac{e^{\frac{i}{2}u(-x)}}{16\sqrt{E}} - \sqrt{E}e^{-\frac{i}{2}u(-x)})\widetilde{\psi}_2(-x,\widetilde{E}) = 0 \qquad 5.1$$

$$\frac{d}{dx}\widetilde{\psi}_2(-x,\widetilde{E}) + \frac{i}{4}\frac{d}{dt}u(-x) \cdot \widetilde{\psi}_2(-x,\widetilde{E}) + (\frac{e^{-\frac{i}{2}u(-x)}}{16\sqrt{E}} - \sqrt{E}e^{\frac{i}{2}u(-x)})\widetilde{\psi}_1(-x,\widetilde{E}) = 0 \qquad 5.2$$

change of variable: $-x \to x$. We find (5) become (4) with eigenvalue $\widetilde{E} = \frac{1}{16^2 E}$. So sectral symmetry in even case:

$$E_j \in \sum^{(b)} \Rightarrow \frac{1}{16^2 E_j} \in \sum^{(b)} \qquad 3$$

has been proved.

**End of proof**.

**Case 2**:

If $u$ is odd functions( p.s. $u(x) = -u(-x)$, kink):

(*pf*) : the same $\begin{pmatrix}\psi_1(x,E)\\ \psi_2(x,E)\end{pmatrix}$ and $E$ are eigenfunction and value of (4); then

$\begin{pmatrix}\psi_1(-x,E)\\ \psi_2(-x,E)\end{pmatrix} = \begin{pmatrix}\widehat{\psi}_1(x,\widetilde{E})\\ \widehat{\psi}_2(x,\widetilde{E})\end{pmatrix}$ where $\widetilde{E} = \frac{1}{16^2 E}$ are eigenfunction and value of (4).

Put above assumation into (4), we get:

$$\frac{d}{d(-x)}\widehat{\psi}_2(-x,\widetilde{E}) - \frac{i}{4}\frac{d}{dt}u(-x) \cdot \widehat{\psi}_2(-x,\widetilde{E}) + (\sqrt{\widetilde{E}}e^{-\frac{i}{2}u(-x)} - \frac{1}{16\sqrt{\widetilde{E}}}e^{\frac{i}{2}u(-x)})\widehat{\psi}_1(-x,\widetilde{E}) = 0 \qquad 6.1$$

$$-\frac{d}{d(-x)}\widehat{\psi}_1(-x,\widetilde{E}) - \frac{i}{4}\frac{d}{dt}u(-x) \cdot \widehat{\psi}_1(-x,\widetilde{E}) + (\sqrt{\widetilde{E}}e^{\frac{i}{2}u(-x)} - \frac{1}{16\sqrt{\widetilde{E}}}e^{-\frac{i}{2}u(-x)})\widehat{\psi}_2(-x,\widetilde{E}) = 0 \qquad 6.2$$

The spectral symmetry in odd case

$$E_j \in \sum^{(k)} \Rightarrow \frac{1}{16^2 E_j} \in \sum^{(k)} \qquad (7)$$

also been proved.

**End of proof**.

Remark: We can shift $x_0$ ( p.s. which means $u(x - x_0) = u(-x - x_0)$, etc.), if $x_0$ is real, the result above is obviously not change; if $x_0$ is imagary, we find we also have spectral symmetry. This has applications because $isc(x+iK';k) = \frac{1}{dn(x;k)}$ if we let $x_0 = iK'$. $sc(x;k)$ is related to kink and is a odd function, while $\frac{1}{dn(x;k)}$ is related to breather and is an even function.

The relation between $\Delta(E)$ and $\Delta(\frac{1}{16^2E})$ is shown in Appendix A. The conclusion is both even and odd caseshas the relation: $\Delta(\frac{1}{16^2E}) = (-1)^M\Delta(E)$; where $u(L) - u(0) = 2\pi M$. Like other discriminant, The Floquet discriminant $\Delta(E)$ give a judgement wherther the state is bloch or not.

### 1.2. Spectral symmetry⇒separable function in $N = 2$:

In [10]~[13], we have study the fundamental theorem related reduction Riemann theta function of higher genus to lower genus theta function. These related to reduction hyperelliptic integrals to elliptic integrals and get the separable solutions. They use involution symmetry for example, but this actually related to spectral symmetry in our case. Here, we only analysis breather case and obtain totally two kinds of N=2 separable solutions in [1].The spectral symmetry is:

$$dI(E) = -16dJ(\epsilon)$$
$$-16dJ(E) = dI(\epsilon) \tag{8}$$

Remark: If SGE has spectral symmetry in $N = 2$, then the only breather solutions are **Case 2.6 (a)**, **(b)** of [**1**].

Because In $N = 2$ spectrum has spectral symmetry, which means $\sum = \{E_1, E_2, E_3 = \frac{1}{16^2E_2}, E_4 = \frac{1}{16^2E_1},\}$. From [**2**]; one of the conditions for real breather in $N = 2$ is we must have two pairs of conjugate spectrum. So negelet the degenerate spectrum case:

**Case 1**: If $E_2 = E_1^*$, then we have **Case (a)**:
$$\sum\nolimits_{breather}^{(a)} = \{E_1, E_2, E_3, E_4 | E_1 = E_2^* = re^{i\phi}, E_3 = E_4^* = \frac{1}{16^2r}e^{i\phi}, 0 < r < \frac{1}{16}, 0 < \phi < \pi\}$$

**Case 2**: If $\frac{1}{16^2E_1} = E_1^*$, then we have **Case (b)**:
$$\sum\nolimits_{breather}^{(b)} = \{E_1, E_2, E_3, E_4 | E_1 = E_4^* = \frac{1}{16}e^{i\phi_1}, E_2 = E_3^* = \frac{1}{16}e^{i\phi_2}, 0 < \phi_1 < \phi_2 < \pi\}$$

**Case 3**: If $\frac{1}{16^2E_2} = E_1^*$, then we have **Case (a)**

So we total have two kinds of N=2 separable solutions, and each breather also have their corresponding kink.

## 2. General formula related to Riemann theta function:

The $\Theta$- representaion, $N = 2$ solutions is $q_{N=2} = 2i\ln[\frac{\Theta(\vec{l}+\frac{\vec{1}}{2};\overleftrightarrow{B})}{\Theta(\vec{l};\overleftrightarrow{B})}]$. where $\Theta(\vec{l};\overleftrightarrow{B}) \equiv \sum_{\vec{k}\in Z\times Z} \exp\{i\pi(<\overleftrightarrow{B}\vec{k},\vec{k}> + 2<\vec{l},\vec{k}>)\}$. We can determine $\vec{l}$ and $\overleftrightarrow{B}$ in term of elliptic integrals of $a_i$ and $b_i$ loop.

(1). We choose the combination of hyperelliptic integrals:

$$dU_1 = \frac{C_{11}z + C_{12}}{\sqrt{P_5(z)}}dz \tag{9}$$

$$dU_2 = \frac{C_{21}z + C_{22}}{\sqrt{P_5(z)}}dz$$

where $P_5(z)$ are the polynomial of fifth order. We normalize them,

$$\begin{pmatrix} \oint_{a_1} dU_1 & \oint_{a_1} dU_2 \\ \oint_{a_2} dU_1 & \oint_{a_2} dU_2 \end{pmatrix} = \begin{pmatrix} \oint_{a_1} \frac{C_{11}z+C_{12}}{\sqrt{P_5(z)}} dz & \oint_{a_1} \frac{C_{21}z+C_{22}}{\sqrt{P_5(z)}} \\ \oint_{a_2} \frac{C_{11}z+C_{12}}{\sqrt{P_5(z)}} dz & \oint_{a_2} \frac{C_{21}z+C_{22}}{\sqrt{P_5(z)}} \end{pmatrix} = \begin{pmatrix} 1 & 0 \\ 0 & 1 \end{pmatrix} \quad (10)$$

Define $dJ = \frac{zdz}{\sqrt{P_5(z)}}$; $dI = \frac{dz}{\sqrt{P_5(z)}}$

(2). Define:

$$\overleftrightarrow{l} = iC \cdot [I(A)]^{-1,t} = iC \cdot \begin{pmatrix} C_{11} & C_{12} \\ C_{21} & C_{22} \end{pmatrix} = iC \cdot \begin{pmatrix} J(a_1) & J(a_2) \\ I(a_1) & I(a_2) \end{pmatrix}^{-1} \quad (11)$$

$C$ is a constant. $I(A) = \begin{pmatrix} J(a_1) & I(a_1) \\ J(a_2) & I(a_2) \end{pmatrix}$; where $I(a_1) = \oint_{a_1} dI$, etc. and superscript $t$ means transport. Also require the form of $I(B)$ must the same with $I(A)$, except $a_i$ loop transform to $b_i$ loop. Then $I(B) = \begin{pmatrix} J(b_1) & I(b_1) \\ J(b_2) & I(b_2) \end{pmatrix}$, define:

$$\overleftrightarrow{B} = I(B) \cdot I(A)^{-1} = \begin{pmatrix} J(b_1) & I(b_1) \\ J(b_2) & I(b_2) \end{pmatrix} \cdot \begin{pmatrix} C_{11} & C_{21} \\ C_{12} & C_{22} \end{pmatrix} = \begin{pmatrix} \oint_{b_1} \frac{C_{11}z+C_{12}}{\sqrt{P_5(z)}} dz & \oint_{b_1} \frac{C_{21}z+C_{22}}{\sqrt{P_5(z)}} \\ \oint_{b_2} \frac{C_{11}z+C_{12}}{\sqrt{P_5(z)}} dz & \oint_{b_2} \frac{C_{21}z+C_{22}}{\sqrt{P_5(z)}} \end{pmatrix} \quad (12)$$

(10) and (11) are N=2 generalized formulas of N=1 list in [1].

(3). Use x- and t-flow basis:

In [4], the author define x-flow $X: \frac{(1+16E)dE}{R(E)}$; t-flow $T: \frac{(1-16E)dE}{R(E)}$, where $R^2$

$(E) = E \prod_{j=1}^{2} (E - E_j)(E - \frac{1}{16^2 E_j})$. Using these basis, we can reduce the hyperelliptic integrals to elliptic integrals and by the separability theorem of [10]~[13], the Riemann theta function of genus=2 in the solution can factor into products of theta functions of genus=1. In new basis,

$$\overleftrightarrow{l} = iC \cdot I(\bar{A})^{-1,t} = iC \cdot \begin{pmatrix} -X(a_1) & T(a_1) \\ -X(a_2) & T(a_2) \end{pmatrix}^{-1,t}$$

$$= iC \cdot \begin{pmatrix} -\oint_{a_1} \frac{(1+16E)dE}{R(E)} & \oint_{a_1} \frac{(1-16E)dE}{R(E)} \\ -\oint_{a_2} \frac{(1+16E)dE}{R(E)} & \oint_{a_2} \frac{(1-16E)dE}{R(E)} \end{pmatrix}^{-1,t}$$

$$= \frac{iC}{32w} \begin{pmatrix} I(a_2) - 16J(a_2) & I(a_2) + 16J(a_2) \\ -I(a_1) + 16J(a_1) & -I(a_1) - 16J(a_1) \end{pmatrix} \quad (13)$$

where $w = \det \begin{pmatrix} I(a_1) & J(a_1) \\ I(a_2) & J(a_2) \end{pmatrix}$. If $\epsilon = \frac{1}{16^2 E}$, we have x-flow antisymmetry and t-flow symmetry:

$$\frac{(1+16E)dE}{R(E)} = -\frac{(1+16\epsilon)d\epsilon}{R(\epsilon)}$$
$$\frac{(1-16E)dE}{R(E)} = \frac{(1-16\epsilon)d\epsilon}{R(\epsilon)} \tag{14}$$

Because we have spectral symmetry.

$$I(\overline{B}) = \begin{pmatrix} -X(b_1) & T(b_1) \\ -X(b_2) & T(b_2) \end{pmatrix} = \begin{pmatrix} -\oint_{b_1} \frac{(1+16E)dE}{R(E)} & \oint_{b_1} \frac{(1-16E)dE}{R(E)} \\ -\oint_{b_2} \frac{(1+16E)dE}{R(E)} & \oint_{b_2} \frac{(1-16E)dE}{R(E)} \end{pmatrix}$$
$$= \begin{pmatrix} -I(b_1) - 16J(b_1) & I(b_1) - 16J(b_1) \\ -I(b_2) - 16J(b_2) & I(b_2) - 16J(b_2) \end{pmatrix} \tag{15}$$

$$\overleftrightarrow{B} = I(\overline{B}) \cdot I(\overline{A})^{-1} = \frac{1}{w}\begin{pmatrix} I(b_1)J(a_2) - I(a_2)J(b_1) & I(a_1)J(b_1) - I(b_1)J(a_1) \\ I(b_2)J(a_2) - I(a_2)J(b_2) & I(a_1)J(b_2) - I(b_2)J(a_1) \end{pmatrix} \tag{16}$$

Although $\overleftrightarrow{B}$ in (16) is not explictly symmetry by the fundenmental property of periods, but we will find this automatlly including in the spectral symmetry.

## 3.Drive two cases of N=2 separable solutions from spectrum symmetry and loop diagrams:

### 3.1 The relation between two cases:

We consider the kink cases, that means the discrete spectrum $\sum_1 = \{\frac{1}{16^2E_1}, \frac{1}{16^2E_2}, E_2, E_1 | \frac{1}{16^2E_1} < \frac{1}{16^2E_2} < E_2 < E_1 < 0\}$. And we define $a_i$ loop are the branch cuts connection with two different discrete spectras and each $a_i$ connect different discrete spectras. Define $b_i$ loop are the branch cuts connect one of the branch points to the radiation branch cuts, (p.s. continuation spectrum $\sum_2 = \{0, \infty\}$.

1.Case(a):

In this case, we define $a_1$ connect $E_1$ to $E_2$ and $a_2$ connect $\frac{1}{16^2E_2}$ to $\frac{1}{16^2E_1}$. From x- and t-flow integrals, we obtain:

$$I(a_1) + 16J(a_1) = -16\oint_A \frac{dz}{\sqrt{2(z-\frac{1}{16})(z-z_1)(z-z_2)}} = w^+$$

$$I(b_1) + 16J(b_1) = -16\oint_B \frac{dz}{\sqrt{2(z-\frac{1}{16})(z-z_1)(z-z_2)}} = w^{+\prime}$$

$$I(a_1) - 16J(a_1) = -16\oint_A \frac{dz}{\sqrt{2(z+\frac{1}{16})(z-z_1)(z-z_2)}} = w^-$$

$$I(b_1) - 16J(b_1) = -16\oint_B \frac{dz}{\sqrt{2(z+\frac{1}{16})(z-z_1)(z-z_2)}} = w^{-\prime} \tag{17}$$

where $z_i = \frac{1}{2}(E_i + \frac{1}{16^2E_i}); i = 1, 2; z = \frac{1}{2}(E + \frac{1}{16^2E})$. From (8), we have the spectrum symmetry, (p.s. we study it in detail in Section 4)

$$I(a_1) = 16J(a_2); I(a_2) = 16J(a_1)$$
$$I(b_1) = 16J(b_2); I(b_2) = 16J(b_1) \tag{18}$$

So $w = \det\begin{pmatrix} I(a_1) & J(a_1) \\ I(a_2) & J(a_2) \end{pmatrix} = \frac{1}{16} w^+ \cdot w^-$ and

$$l = \frac{iC}{32w}\begin{pmatrix} I(a_2) - 16J(a_2) & I(a_2) + 16J(a_2) \\ -I(a_1) + 16J(a_1) & -I(a_1) - 16J(a_1) \end{pmatrix} = \frac{iC}{2}\begin{pmatrix} -\frac{1}{w^+} & \frac{1}{w^-} \\ -\frac{1}{w^+} & -\frac{1}{w^-} \end{pmatrix} \quad (19)$$

From (16)

$$\overleftrightarrow{B} = \frac{1}{2}\begin{pmatrix} \tau^+ + \tau^- & \tau^+ - \tau^- \\ \tau^+ - \tau^- & \tau^+ + \tau^- \end{pmatrix} \quad (20)$$

where $\tau^+ = \frac{w^{+\prime}}{w^+}$ and $\tau^- = \frac{w^{-\prime}}{w^-}$. By [1], two separable modular parameter of Jacobi elliptic function are $\tau_1 = 2\tau^+$ and $\tau_2 = 2\tau^-$. Actually, this case also corresponding to Case 1 in section 1.2 in breather case. Also Case 3 there defining $a_1$ connect with $E_1$ to $\frac{1}{16^2 E_2}$ and $a_2$ connect $E_2$ to $\frac{1}{16^2 E_1}$ relatively in kink. This is also including in Case (a) because of the same relations, (18).

2. Case(b):

In this case, we define $a_1$ connect $E_1$ to $\frac{1}{16^2 E_1}$ and $a_2$ connect $E_2$ to $\frac{1}{16^2 E_2}$. This case is special enough so that we even cannot calculate by x- and t-flow integrals in (17) because the shrink range of $A$-loop. But from (8) and the loop diagrams, we find: (p.s. we also discuss in detail in Section 4):

$$I(a_2) = -16J(a_2); \quad I(a_1 + a_2) = 16J(a_1 + a_2) \quad (21)$$
$$I(b_1) = 16J(b_1); \quad I(b_2 - b_1) = -16J(b_2 - b_1)$$

We rearrange (21)

$$I(a_1) = 16J(a_1 + 2a_2); \quad I(a_1 + 2a_2) = 16J(a_1)$$
$$I(2b_1 - b_2) = 16J(b_2); \quad I(b_2) = 16J(2b_1 - b_2) \quad (22)$$

If we let

$$a'_{1b} = a_{1b}; \quad a'_{2b} = a_{1b} + 2a_{2b}; \quad b'_{1b} = 2b_{1b} - b_{2b}; \quad b'_{2b} = b_{2b} \quad (23)$$

So that (22) is similar to (18), (p.s. change loop after spectral exchange). Also lower index $b$ in the right hand side indicate the original loops in Case(b). If we change loops

$$b'_i = \sum_j a_{ij} \cdot b_j + \sum_j b_{ij} \cdot a_j$$
$$a'_i = \sum_j c_{ij} \cdot b_j + \sum_j d_{ij} \cdot a_j \quad (24)$$

Then we say $\overleftrightarrow{B}'$ and $\overleftrightarrow{B}$ are equivalent if $\overleftrightarrow{B}' = \overleftrightarrow{\sigma} \cdot \overleftrightarrow{B}$. (p.s. $\overleftrightarrow{B}' = [(a_{ij}) \cdot \overleftrightarrow{B} + (b_{ij})] \cdot [(c_{ij}) \cdot \overleftrightarrow{B} + (d_{ij})]^{-1}$). Where $\overleftrightarrow{\sigma} = \begin{pmatrix} a_{ij} & b_{ij} \\ c_{ij} & d_{ij} \end{pmatrix} \in Sp(2, Z)$ and $(a_{ij}) \sim (d_{ij})$ are $2 \times 2$ integer matrix. Also require $\det\begin{pmatrix} a_{ij} & b_{ij} \\ c_{ij} & d_{ij} \end{pmatrix} = 1$ and

$\begin{pmatrix} a_{ij} & b_{ij} \\ c_{ij} & d_{ij} \end{pmatrix} \cdot \begin{pmatrix} 0 & 1 \\ -1 & 0 \end{pmatrix} \cdot \begin{pmatrix} a^t_{ij} & c^t_{ij} \\ b^t_{ij} & d^t_{ij} \end{pmatrix} = \begin{pmatrix} 0 & 1 \\ -1 & 0 \end{pmatrix}$. where $a^t_{ij}$ means the transport of $a_{ij}$. From (23), we get:

$$(a_{ij}) = \begin{pmatrix} 2 & -1 \\ 0 & 1 \end{pmatrix}; (b_{ij}) = (c_{ij}) = \begin{pmatrix} 0 & 0 \\ 0 & 0 \end{pmatrix}; (d_{ij}) = \begin{pmatrix} 1 & 0 \\ 1 & 2 \end{pmatrix} \qquad (25)$$

We obtain

$$\begin{pmatrix} 2 & -1 \\ 0 & 1 \end{pmatrix} \cdot \begin{pmatrix} \alpha & \alpha \\ \alpha & \beta \end{pmatrix} \cdot \begin{pmatrix} 1 & 0 \\ 1 & 2 \end{pmatrix}^{-1} = \begin{pmatrix} \frac{\beta}{2} & \alpha - \frac{\beta}{2} \\ \alpha - \frac{\beta}{2} & \frac{\beta}{2} \end{pmatrix}$$

$$\Rightarrow \begin{pmatrix} 1 & 1 \\ 0 & 2 \end{pmatrix} \cdot \begin{pmatrix} \frac{\beta}{2} & \alpha - \frac{\beta}{2} \\ \alpha - \frac{\beta}{2} & \frac{\beta}{2} \end{pmatrix} \cdot \begin{pmatrix} 2 & 0 \\ -1 & 1 \end{pmatrix}^{-1} = \begin{pmatrix} \alpha & \alpha \\ \alpha & \beta \end{pmatrix}$$

Later (p.s. (41)~(44)) we will prove $\begin{pmatrix} \alpha & \alpha \\ \alpha & \beta \end{pmatrix}$ is the period of Case(b), while we have known $\begin{pmatrix} \frac{\beta}{2} & \alpha - \frac{\beta}{2} \\ \alpha - \frac{\beta}{2} & \frac{\beta}{2} \end{pmatrix}$ is the period of Case(a). Because $\det \begin{pmatrix} a_{ij} & b_{ij} \\ c_{ij} & d_{ij} \end{pmatrix} = 4 \neq 1$ and

$$\begin{pmatrix} a_{ij} & b_{ij} \\ c_{ij} & d_{ij} \end{pmatrix} \cdot \begin{pmatrix} 0 & 1 \\ -1 & 0 \end{pmatrix} \cdot \begin{pmatrix} a^t_{ij} & c^t_{ij} \\ b^t_{ij} & d^t_{ij} \end{pmatrix} = \begin{pmatrix} 0 & 2 \\ -2 & 0 \end{pmatrix} \neq \begin{pmatrix} 0 & 1 \\ -1 & 0 \end{pmatrix}.$$ Case(a) and Case(b) are not equivalent but are Landen transformation between them. ( the final part in this section in detail). From (26), we have another basis depend on Case(a):

$$b'_{1a} = b_{1a} + b_{2a}; \ b'_{2a} = 2b_{2a}; \ a'_{1a} = 2a_{1a}; \ a'_{2a} = -a_{1a} + a_{2a} \qquad (27)$$

Where lower index $a$ on the right hand side represent original loops of Case(a). While by (26) the left hand side represent loops of Case(b). Using (27), we can calculate $\overleftrightarrow{l}$ and $\overleftrightarrow{B}$ of Case(b) from the result of Case(a).

$$\overleftrightarrow{l}' = \frac{iC}{32w'} \begin{pmatrix} I(a'_{2a}) - 16J(a'_{2a}) & I(a'_{2a}) + 16J(a'_{2a}) \\ -I(a'_{1a}) + 16J(a'_{1a}) & -I(a'_{1a}) - 16J(a'_{1a}) \end{pmatrix} = \frac{iC}{2} \begin{pmatrix} -\frac{1}{w^+} & 0 \\ -\frac{1}{w^+} & -\frac{1}{w^-} \end{pmatrix} \qquad (28)$$

where $w' \det \begin{pmatrix} I(a'_{1a}) & J(a'_{1a}) \\ I(a'_{2a}) & J(a'_{2a}) \end{pmatrix} = \frac{1}{8} w^+ w^-$.

$$\overleftrightarrow{B}' = \frac{1}{32w'} \begin{pmatrix} -I(b'_{1a}) - 16J(b'_{1a}) & I(b'_{1a}) - 16J(b'_{1a}) \\ -I(b'_{2a}) - 16J(b'_{2a}) & I(b'_{2a}) - 16J(b'_{2a}) \end{pmatrix} \cdot \begin{pmatrix} I(a'_{2a}) - 16J(a'_{2a}) & -I(a'_{1a}) + 16J(a'_{1a}) \\ I(a'_{2a}) + 16J(a'_{2a}) & -I(a'_{1a}) - 16J(a'_{1a}) \end{pmatrix}$$

$$= \begin{pmatrix} \tau^+ & \tau^+ \\ \tau^+ & \tau^+ + \tau^- \end{pmatrix} \qquad (29)$$

This prove the statement below (26). The two one dimensional modular parameter of $\overleftrightarrow{B}_a$ are $\tau_1 = 2\tau^+$ and $\tau_2 = 2\tau^-$. And the two one dimensional modular parameter of $\overleftrightarrow{B}_a$ are $\tau_3 = 4\tau^+$ and $\tau_4 = -\tau^-$. where $\overleftrightarrow{B}_b = \begin{pmatrix} \tau^+ & \tau^+ \\ \tau^+ & \tau^+ + \tau^- \end{pmatrix}$ and

$\overleftrightarrow{B}_a = \frac{1}{2} \begin{pmatrix} \tau^+ + \tau^- & \tau^+ - \tau^- \\ \tau^+ - \tau^- & \tau^+ + \tau^- \end{pmatrix}$ From [11], we can choose $\overleftrightarrow{\sigma}_a = \begin{pmatrix} 0 & -1 & 0 & 0 \\ 0 & 0 & 1 & 0 \\ 0 & 0 & 1 & -1 \\ -1 & -1 & 0 & 0 \end{pmatrix}$ so

that from (20), $\overleftrightarrow{\sigma}_a \cdot \overleftrightarrow{B}_a = \begin{pmatrix} \frac{\tau^-}{2} & \frac{1}{2} \\ \frac{1}{2} & -\frac{1}{2\tau^+} \end{pmatrix}$.

### 3.2 Further explain the relation between two cases

In [11], the Riemann theta function of the first order with characteristic $[\alpha; \beta]$, $\alpha, \beta \in R^g$, is the function

$$\Theta[\alpha;\beta](\vec{l};\overleftrightarrow{B}) \equiv \sum_{\vec{k} \in Z^g} \exp\{i\pi(<\overleftrightarrow{B}(\vec{k}+\vec{\alpha}),(\vec{k}+\vec{\alpha})> +2 <\vec{l}+\vec{\beta},\vec{k}+\vec{\alpha}>)\} \quad (30)$$

If $g = 2$, then we have the relation

$$\Theta[\alpha;\beta](\vec{l}; \begin{pmatrix} B_{11} & 0 \\ 0 & B_{22} \end{pmatrix}) = e^{2\pi i \alpha_1 \alpha_2} \Theta[\alpha_1,\alpha_2;\beta_1-\alpha_2,\beta_2-\alpha_1](\vec{l}; \begin{pmatrix} B_{11} & 1 \\ 1 & B_{22} \end{pmatrix}) \quad (31)$$

We can vanish the off diagonal term by doing the second order transformation, following by (31). Using the symbols of $\tau_1$ and $\tau_2$. Then $\overleftrightarrow{\sigma}_a \cdot \overleftrightarrow{B}_a \to \begin{pmatrix} \frac{\tau_2}{2} & 0 \\ 0 & -\frac{2}{\tau_1} \end{pmatrix}$. If we call the combine matrix of (25) is $\sigma_c$ ( p.s. $\sigma_c = \begin{pmatrix} 2 & -1 & 0 & 0 \\ 0 & 1 & 0 & 0 \\ 0 & 0 & 1 & 0 \\ 0 & 0 & 1 & 2 \end{pmatrix}$ ), then the relation of (26) is simplily $\sigma_c \cdot \overleftrightarrow{B}_b = \overleftrightarrow{B}_a$. Now we want choose $\sigma_b = \begin{pmatrix} 0 & -1 & 0 & 0 \\ 0 & 0 & 1 & 0 \\ 0 & 0 & 0 & -2 \\ -2 & 0 & 0 & 0 \end{pmatrix}$, ( p.s. see Appendix B) then $\sigma_b \cdot \overleftrightarrow{B}_b = \overleftrightarrow{\sigma}_a \cdot \overleftrightarrow{B}_a = \begin{pmatrix} \frac{\tau^-}{2} & \frac{1}{2} \\ \frac{1}{2} & -\frac{1}{2\tau^+} \end{pmatrix}$. Doing the same process as case(a) to vanish the off diagonal term and using the symbols of $\tau_3$ and $\tau_4$. Then

$\sigma_b \cdot \overleftrightarrow{B}_b \to \begin{pmatrix} -\tau_4 & 0 \\ 0 & -\frac{4}{\tau_3} \end{pmatrix}$. We find the diagonal term from $\frac{\tau_2}{2}$ to $-\tau_4$, while $-\frac{2}{\tau_1}$ to $-\frac{4}{\tau_3}$.

From (23) or (27), we change and double the loops of their periods. We have "Modular and Landen transformation" between these two cases after the above surgery. We can write down them explicitly to show the transformations explicitly.

## 4. Explict list these two solution and discuss:

Here we use the symbols of [1]:

### 4.1 Case(a):

[𝔸]: **Breather**:

$$\sum\nolimits_{breather}^{(a)} = \{E_1, E_2, E_3, E_4 | E_1 = E_2^* = re^{i\phi}, E_3 = E_4^* = \frac{1}{16^2 r} e^{i\phi}, 0 < r < \frac{1}{16}, 0 < \phi < \pi\} \quad (32)$$

From (19) and (20) and the require of breather in [2]., we obtain

$$\vec{l}^{(a)}_{breather} = \frac{iC}{2}\begin{pmatrix} -\frac{1}{w^+}x + \frac{1}{w^-}t \\ -\frac{1}{w^+}x - \frac{1}{w^-}t \end{pmatrix}; \overleftrightarrow{B}^{(a)}_{breather} = \begin{pmatrix} \frac{1}{2} & 0 \\ 0 & \frac{1}{2} \end{pmatrix} + \frac{1}{2}\begin{pmatrix} \tau^+ + \tau^- & \tau^+ - \tau^- \\ \tau^+ - \tau^- & \tau^+ + \tau^- \end{pmatrix}.$$ $\vec{l}$ and $\overleftrightarrow{B}$ are defined in Riemann theta function in section 2.

Here we let $\vec{l}^{(a)}_{breather}$ and $\overleftrightarrow{B}^{(a)}_{breather}$ as $\vec{l}^{(a)}_b$ and $\vec{B}^{(a)}_b$ for shorthand:

$$\Theta^{(a)}_{breather}(\vec{l}^{(a)}_b, \overleftrightarrow{B}^{(a)}_b) = \theta_4(\frac{iC}{w^+} \cdot x; 2\tau^+) \cdot \theta_4(\frac{iC}{w^-} \cdot t; 2\tau^-) + i\theta_2(\frac{iC}{w^+} \cdot x; 2\tau^+) \cdot \theta_2(\frac{iC}{w^-} \cdot t; 2\tau^-) \quad (33)$$

and because

$$\Theta(\vec{l} + \frac{\vec{1}}{2}, \overleftrightarrow{B}) = \Theta(\vec{l}, \overleftrightarrow{B})^* \tag{35}$$

So in this case

$$q_2(x, t, \sum^{(a)}_{breather}) = 2i\ln[\frac{\Theta(\vec{l} + \frac{\vec{1}}{2}, \overleftrightarrow{B})}{\Theta(\vec{l}, \overleftrightarrow{B})}] = 4\tan^{-1}[\frac{\text{Im}\,\Theta(\vec{l}, \overleftrightarrow{B})}{\text{Re}\,\Theta(\vec{l}, \overleftrightarrow{B})}]$$

$$= 4\tan^{-1}[\sqrt{\frac{k(2\tau^+)k(2\tau^-)}{k'(2\tau^+)k'(2\tau^-)}} \cdot nc(\frac{2C}{w^+}K(2\tau^+) \cdot x; k'(2\tau^+)) \cdot nc(\frac{2C}{w^-}K(2\tau^-) \cdot t; k'(2\tau^-))]$$

[𝔹]: **The corresponding kink**:

$$\sum^{(a)}_{kink} = \{E_1, E_2, E_3, E_4 | E_1 = -re^{-\eta}, E_2 = -re^{\eta}, E_3 = \frac{1}{16^2 E_2} = \frac{-1}{16^2 r}e^{-\eta},$$

$$E_4 = \frac{1}{16^2 E_1} = \frac{-1}{16^2 r}e^{\eta}, 0 < r < \frac{1}{16}\} \tag{37}$$

here we instead $\phi$ in (32) to get the kink spectrum, (37) by:

$$\phi = \pi + i\eta \tag{38}$$

In this case $\vec{l}^{(a)}_{kink} = \begin{pmatrix} \frac{1}{4} \\ \frac{1}{4} \end{pmatrix} + \frac{iC}{2}\begin{pmatrix} -\frac{1}{w^+}x + \frac{1}{w^-}t \\ -\frac{1}{w^+}x - \frac{1}{w^-}t \end{pmatrix}; \overleftrightarrow{B}^{(a)}_{kink} = \frac{1}{2}\begin{pmatrix} \tau^+ + \tau^- & \tau^+ - \tau^- \\ \tau^+ - \tau^- & \tau^+ + \tau^- \end{pmatrix}.$

Because $\frac{1}{16^2 E_1} < \frac{1}{16^2 E_2} < E_2 < E_1 < 0$; The $a_i$ and $b_i$ loop are defined in Case(a) of Section 3. Here we want to focus on $b_1$ loop. We have defined in the begining of Section 3. $b_1$ loop is connected with one of the branch ones in $a_1$, $E_1$ or $E_2$ to one of the radiation branch cut, 0 or ∞. The similar definition is also for $b_2$ loop. Finally, we also need to compacitify ∞ and −∞.

$$I(b_1) = \oint_{b_1} dI(E) = 2\int_{E_1}^{0} dI(E) = 32\int_{\infty}^{\frac{1}{16^2 E_1}} dJ(\epsilon = \frac{1}{16^2 E}) = 16J(b_2)$$

$$I(b_2) = \oint_{b_2} dI(E) = 2\int_{\frac{1}{16^2 E_1}}^{0} dI(E) = 32\int_{-\infty}^{E_1} dJ(\epsilon = \frac{1}{16^2 E}) = 16J(b_1) \tag{39}$$

We get the Riemann theta function in this case: We let $\vec{l}^{(a)}_{kink}$ and $\overleftrightarrow{B}^{(a)}_{kink}$ as $\vec{l}^{(a)}_k$ and $\vec{B}^{(a)}_k$ for shorthand:

$$\Theta^{(a)}_{kink}(\vec{l}^{(a)}_k, \overleftrightarrow{B}^{(a)}_k) = \theta_4(\frac{iC}{w^+} \cdot x; 2\tau^+) \cdot \theta_3(\frac{iC}{w^-} \cdot t; 2\tau^-) - \theta_1(\frac{iC}{w^+} \cdot x; 2\tau^+) \cdot \theta_2(\frac{iC}{w^-} \cdot t; 2\tau^-)$$

$$\Theta^{(a)}_{kink}(\vec{l}^{(a)}_k + \frac{\vec{1}}{2}, \overleftrightarrow{B}^{(a)}_k) = \theta_4(\frac{iC}{w^+} \cdot x; 2\tau^+) \cdot \theta_3(\frac{iC}{w^-} \cdot t; 2\tau^-) + \theta_1(\frac{iC}{w^+} \cdot x; 2\tau^+) \cdot \theta_2(\frac{iC}{w^-} \cdot t; 2\tau^-) \tag{40}$$

So the corresponding kink is:

$$q_2(x,t,\sum_{kink}^{(a)}) = 2i\ln[\frac{\Theta(\vec{l}+\frac{\vec{1}}{2},\overleftrightarrow{B})}{\Theta(\vec{l},\overleftrightarrow{B})}]$$

$$= 4\tan^{-1}[\sqrt{k(2\tau^+)k(2\tau^-)} \cdot sc(\frac{2C}{w^+} \cdot K(2\tau^+) \cdot x; k'(2\tau^+)) \cdot nd(\frac{2C}{w^-} \cdot K(2\tau^-) \cdot t; k'(2\tau^-))]$$

## 4.2 Case(b):

[𝔸]: **The kink**:

$$\sum_{kink}^{(b)} = \{E_1, E_2, E_3, E_4 | E_1 = \frac{-1}{16}e^{-\eta_1}, E_2 = \frac{-1}{16}e^{-\eta_2}, E_3 = \frac{1}{16^2 E_2} = \frac{-1}{16}e^{\eta_2},$$

$$E_4 = \frac{1}{16^2 E_1} = \frac{-1}{16}e^{\eta_1}; \eta_1 > \eta_2 > 0\} \tag{42}$$

[**1**]. **Explain the relations in (21)**:

The kink spectrum is $E_4 < E_3 < E_2 < E_1 < 0$. We choose $a_1$ loop connect $E_1$ and $E_4$ with clock-wise. $a_2$ loop connect $E_2$ and $E_3$ with reverse clock-wise. By the principle in Section 3, $b_1$ loop connect $E_1$ and 0. We will use the equivalent loop of $b_1$: $\overline{E_1, 0} \sim \overline{E_4, 0} \sim \overline{E_4, \infty} \sim \overline{-\infty, E_4}$ to explain the relations in (21). $b_2$ loop connect $E_2$ and 0. $b_1$ and $b_2$ loop with the same clock-wise.

(1). The first relation come from (8):

$$I(a_2) = \oint_{a_2} dI(E) = 2\int_{E_2}^{E_3} dI(E) = -32\int_{E_3}^{E_2} dJ(\epsilon) = -16J(a_2) \tag{43}$$

The last equality seem change sign because we circle around the same loop: $a_2$. When we transform by (8), it also reverse the direction of clock-wise.

(2). The second relation using the opposite clock-wise of $a_1$ and $a_2$ loop:

$$I(a_1 + a_2) = \oint_{a_1+a_2} dI(E) = 2(\int_{E_2}^{E_1} + \int_{E_4}^{E_3})dI(E) = -32(\int_{E_3}^{E_4} + \int_{E_1}^{E_2})dJ(\epsilon) = 16 * 2(\int_{E_2}^{E_1} + \int_{E_4}^{E_3})dJ(\epsilon) = 16J(a_1 + a_2)$$

(3). The third relation is:

$$I(b_1) = \oint_{b_1} dI(E) = 2\int_{E_1}^{0} dI(E) = -32\int_{\frac{1}{16^2 E_1}=E_4}^{-\infty} dJ(\epsilon) = 16 * 2\int_{-\infty}^{E_4} dJ(\epsilon) = 16J(b_1) \tag{45}$$

This relation different from (43) with opposite sign because we use equivalent loop of $b_1$, not the same loop. Because $b_2 - b_1$ loop connect $E_2$ and $E_1$,

(4). The fourth relation is:

$$I(b_2 - b_1) = \oint_{b_2-b_1} dI(E) = 2\int_{E_2}^{E_1} dI(E) = -32\int_{\frac{1}{16^2 E_2}=E_3}^{\frac{1}{16^2 E_1}=E_4} dJ(\epsilon) = -16 * 2(\int_{E_3}^{0} - \int_{E_4}^{0})dJ(\epsilon) = -16J(b_2 - b_1)$$

The last equiality also use the equivalent loop of $b_2$: $\overline{E_2, 0} \sim \overline{E_3, 0} \sim \overline{E_3, \infty} \sim \overline{-\infty, E_3}$, so the equivalent loop of $b_2 - b_1$: $\overline{E_2, E_1} \sim \overline{E_3, E_4}$.

[**2**]. : **List the kink solution in this case of SGE**:

Using (21), we get $\vec{l}_k^{(b)} = \begin{pmatrix} \frac{1}{4} \\ \frac{1}{4} \end{pmatrix} + \frac{iC}{2}\begin{pmatrix} -\frac{1}{w^+}x \\ -\frac{1}{w^+}x - \frac{1}{w^-}t \end{pmatrix}$; $\overleftrightarrow{B}_k^{(b)} = \begin{pmatrix} \tau^+ & \tau^+ \\ \tau^+ & \tau^+ + \tau^- \end{pmatrix}$. The

$Re(\vec{l}) = \frac{1}{4}$ come from the require of kink in [2].

$\vec{l}_k^{(b)}$ and $\overleftrightarrow{B}_k^{(b)}$ give us the Riemann theta function in this case:

$$\Theta_k^{(b)}(\vec{l}_k^{(b)}; \overleftrightarrow{B}_k^{(b)}) = \theta_4(\frac{iC}{w^+} \cdot x; 4\tau_+) \cdot \theta_2(\frac{iC}{2w^-} \cdot t; \tau^-) - \theta_1(\frac{iC}{w^+} \cdot x; 4\tau^+) \cdot \theta_3(\frac{iC}{2w^-} \cdot t; \tau^-)$$

$$\Theta_k^{(b)}(\vec{l}_k^{(b)} + \frac{1}{2}; \overleftrightarrow{B}_k^{(b)}) = \theta_4(\frac{iC}{w^+} \cdot x; 4\tau_+) \cdot \theta_2(\frac{iC}{2w^-} \cdot t; \tau^-) + \theta_1(\frac{iC}{w^+} \cdot x; 4\tau^+) \cdot \theta_3(\frac{iC}{2w^-} \cdot t; \tau^-) \quad (47)$$

The kink in Case(b) is:

$$q_2(x,t, \sum_{kink}^{(b)}) = 2i \ln[\frac{\Theta(\vec{l}_k^{(b)} + \frac{1}{2}, \overleftrightarrow{B}_k^{(b)})}{\Theta(\vec{l}_k^{(b)}, \overleftrightarrow{B}_k^{(b)})}]$$

$$= 4\tan^{-1}[\sqrt{\frac{k(4\tau^+)}{k(4\tau^-)}} \cdot sc(\frac{2C}{w^+} \cdot K(4\tau^+) \cdot x; k'(4\tau^+)) \cdot dn(\frac{C}{w^-} \cdot K(\tau^-) \cdot t; k'(\tau^-))]$$

[$\mathbb{B}$] : **breather**:

[1]. **Relations between elliptic integrals**:

$$\sum_{breather}^{(b)} = \{E_1, E_2, E_3, E_4 | E_1 = E_4^* = \frac{1}{16}e^{i\phi_1}, E_2 = E_3^* = \frac{1}{16}e^{i\phi_2}, 0 < \phi_1 < \phi_2 < \pi\} \quad (49)$$

In this case $\vec{l}_b^{(b)} = \frac{iC}{2}\begin{pmatrix} -\frac{1}{w^+}x \\ -\frac{1}{w^+}x - \frac{1}{w^-}t \end{pmatrix}$; $\overleftrightarrow{B}_b^{(b)} = \begin{pmatrix} \frac{1}{2} & 0 \\ 0 & \frac{1}{2} \end{pmatrix} + \begin{pmatrix} \tau^+ & \tau^+ \\ \tau^+ & \tau^+ + \tau^- \end{pmatrix}$. The

relation between (42) and (49) is: $\phi_i = \pi + i\eta_i$. Also, if we instead of $r$ in (32) by $\frac{1}{16}e^{i\varphi}$ and let $\phi_1 = \phi + \varphi$ and $\phi_2 = \phi - \varphi$, we obtain thespectrum of (49).

[2]: **List the breather solution in this case of SGE**:

With these $\vec{l}$ and $\overleftrightarrow{B}$, we can obtain:

$$\Theta(\vec{l}_b^{(b)}, \overleftrightarrow{B}_b^{(b)}) = \theta_3(\frac{iC}{w^+} \cdot x; 4\tau^+) \cdot \theta_4(\frac{iC}{2w^-} \cdot t; \tau^-) + i\theta_2(\frac{iC}{w^+} \cdot x; 4\tau^+)\theta_3(\frac{iC}{2w^-} \cdot t; \tau^-) \quad (50)$$

So in this case

$$q_2(x,t, \sum_{breather}^{(b)}) = 2i \ln[\frac{\Theta(\vec{l}_b^{(b)} + \frac{1}{2}; \overleftrightarrow{B}^{(b)}b)}{\Theta(\vec{l}_b^{(b)}; \overleftrightarrow{B}_b^{(b)})}] = 4\tan^{-1}[\frac{\text{Im}\,\Theta(\vec{l}, \overleftrightarrow{B})}{\text{Re}\,\Theta(\vec{l}, \overleftrightarrow{B})}]$$

$$= 4\tan^{-1}[\sqrt{\frac{k(4\tau^+)}{k'(\tau^-)}} \cdot nd[\frac{2C}{w^+} \cdot K(4\tau^+) \cdot x; k'(4\tau^+)] \cdot dc[\frac{C}{w^-} \cdot K(\tau^-) \cdot t; k'(\tau^-)]]$$

From (36) and (40), the modular parameters of Case(a) is $2\tau^+$ and $2\tau^-$ while From (41) and (52), the modular parameters of Case(b) is $4\tau^+$ and $\tau^-$. This explain the "Landen transformation" between two cases in Sec.3. Also There are "Modular transformation" between each Jacoi elliptic function.

[3]. **Some question happen at $\overleftrightarrow{B}$ matrix in this case**:

We can see $\text{Re}\,B_{11} = \frac{1}{2} \neq \text{Re}\,B_{21}$ in this case. But $\text{Re}\,B_{11} = \text{Re}\,B_{21}$ for the kink of Case(b). Using (21) into $\overleftrightarrow{B}$, we obtain:

$$B_{11} = B_{12} = B_{21} = \frac{I(a_2)}{w}[J(b_1) + \frac{1}{16}I(b_1)] = \frac{I(b_1)}{I(a_1) + I(a_2)}$$

$$B_{22} = \frac{1}{w}[-I(a_1)J(b_2) + \frac{I(b_2)}{16}(I(a_1) + 2I(a_2))] = \frac{I(b_2)}{I(a_2)} - \frac{I(a_1)I(b_1)}{I(a_2)[I(a_1) + I(a_2)]} \quad (53)$$

This seem contradiction because $B_{11} = B_{12} = B_{21}$ in (53).The contradiction also appear in our previous work, [9], where we discuss in the Section 6 and 7 in phase N=1. The conclusion is $B_{11} = B_{12} = B_{21}$ both in the kink the breather in Case(b). What let

$\operatorname{Re} B_{11} = \frac{1}{2} \neq \operatorname{Re} B_{21}$ in the breather is the replacement: $\phi_i = \pi + i\eta_i$. As we have said in [9], this replacment will total change the property of modular. For example: In [9], $s'_1 = e^{-\eta}$ satisfy the requirement : $-1 \leq s'_1 \leq 1$, but if we replace $\varphi \leftrightarrow \pi + i\eta$, $s'_1 = -e^{i\varphi}$, which is not satisfied by the traditional requirement for modular parameter. This is the result of analytical continuation. Let us discuss it concretely. How to generate $\frac{1}{2} = \operatorname{Re} B_{11} = \operatorname{Re} B_{22}$ and $0 = \operatorname{Re} B_{12} = \operatorname{Re} B_{21}$? The key is when put the following known facts of $b_i$ and $a_i$ loop relations ( p.s. For example, when we calculate $B$ in Table II in Appendix D of [4] )

$$J(b_1) = \frac{1}{2}J(a_1) + \cdots = \frac{1}{2}\frac{1}{16}[I(a_1) + 2I(a_2)] + \cdots$$
$$I(b_1) = \frac{1}{2}I(a_1) + \cdots \qquad (54)$$

If we put (54) in the last second, then we obtain $\frac{1}{2}$ in $\operatorname{Re} B_{11}$ explictly. Otherwise, we can not obtain $\frac{1}{2}$, and related to $B_{12}$ and $B_{21}$. ( p.s. put (54) in the finalof (53)). So put (54) in different ordering give us different answers.

In $B_{22}$, we can use following relations

$$J(b_2) = \frac{1}{2}J(a_2) + \cdots = -\frac{1}{2}\frac{1}{16}I(a_2) + \cdots$$
$$I(b_2) = \frac{1}{2}I(a_2) + \cdots \qquad (55)$$

Put (55) in the second line in (53), no matter how ordering is, we can obtain $\frac{1}{2}$ in $\operatorname{Re} B_{22}$.

## 5.Miscellious topics related to N=2 separable solutions:

### 5.1 A unify of these N=2 separable case

Because $dn(u + K) = k' \cdot nd(u)$. So $q_2(x, t, \sum_{kink}^{(a)}; t_0) = q_2(x, t, \sum_{kink}^{(b)}; t_0 + \frac{K'_2}{4K_2 b})$. Also we can see $\sum_{kink}^{(a)}$ and $\sum_{kink}^{(b)}$ with $\{r, \eta\} \to \{\eta_1, \eta_2\}$. The kink case of two case are " the same" and branch points are the same. The different is how to "construct" branch cuts, but this is man-made in order to calculate loop integrals.

Case 1: (1): branch cuts are disjoint or (2): intersect some part in the negative real axis. (p.s. depend on $r$ and $\eta$). Then this below to the first case.

Case 2: One branch cut are completely inside another branch cut. Then this below to the second case.

So the spectrum ( p.s. the branch points) depend on the way we connect the branch points into branch cuts. Different choose give us different cases. The step of "chose" branch cut is "**spontaneously symmetry broken**".

### 5.2 The static solutions of N=2:

Because we have proven the transformation between the static N=2 soultions and N=1 solutions in the Appendix of [9]. We lovely discuss them further from our solutions.

(1). Static kink of both cases: From (41) and (48), the static condition is let $k'_2 = 0$, so $\tau_2 = \frac{iK'_2}{K_2} = \frac{i\frac{\pi}{2}}{\infty} = 0 \Rightarrow \alpha = \beta$ and both cases give us $\tau_1 = 4i\alpha$ and $\overleftrightarrow{B} = i\begin{pmatrix} \alpha & \alpha \\ \alpha & \alpha \end{pmatrix}$; The static kink is:

$$q_{kink}(x) = 4\tan^{-1}[\sqrt{k'} \cdot sc(\frac{x}{1 - k'}; k)] \qquad (56)$$

(2). Static breather in Case(a): From (36), the static condition is let $k'_2 = \infty$, so

$k_2 = i\infty \Rightarrow \tau_2 = \frac{iK_2'}{K_2} = \lim_{t \to 0} \frac{iK'(\frac{1}{t})}{K(\frac{1}{t})} = \lim_{t \to 0} \frac{iK'(t)}{K(t)+iK'(t)} = 1 \Rightarrow \beta = \alpha + \frac{1}{2i} \Rightarrow \tau_1 = 1 + 4i\alpha$ and

$\overleftrightarrow{B} = \begin{pmatrix} 1 & 0 \\ 0 & 1 \end{pmatrix} + i \begin{pmatrix} \alpha & \alpha \\ \alpha & \alpha \end{pmatrix}$; The static breather of Case(a) is:

$$q^{(a)}_{breather}(x) = 4\tan^{-1}\{\sqrt{\frac{ik'}{k}}\, nc[(k \pm ik')x; k]\} \tag{57}$$

(3). Static breather in Case(b): From (52), the static condition is let $k_2' = 1$, so

$\tau_2 = i\infty \Rightarrow \beta = \infty$ and $\overleftrightarrow{B} = \begin{pmatrix} \frac{1}{2} & 0 \\ 0 & \frac{1}{2} \end{pmatrix} + i \begin{pmatrix} \alpha & \alpha \\ \alpha & \infty \end{pmatrix}$; The static breather of Case(b) is:

$$q^{(b)}_{breather}(x) = 4\tan^{-1}[\sqrt{k'} \cdot nd(\frac{x}{1+k'}; k)] \tag{58}$$

### 5.3 The modular transformation between static condition in Sec. 5.2:

The static conditions in three cases are $\tau_2 = 0, 1, i\infty$, which correspond to $k_2 = 1, 0, \infty$. These are cusp points in elliptic fundanmental region. The discriminant $\Delta = 0$ in these points means we have degenerate roots. So in static condition, we have degererate branch points and originally N=2 case shrink to "equivalent N=1" case. Actually, In Appendix of [9], we have shown the transformation between N=1 and N=2 static solutions. This fit our expectation because original N=2 solution contract to genus=1 solution due to static condition.

In the time part, we have two part of transformation: **[A]**. From (36) to (41), we have the modular transformation: $\widetilde{k}' = \frac{1}{k'}$. and **[B]**. From (48) to (52), we have the modular transformation: $\widehat{k}' = k$. By these two modular transformations, the static conditions can be unify to $\tau_2 = 0$. Recall period $\tau = 0$ means this period decay to zero and the double period elliptic functions are become single period function. So the time part of (36), (41), (48), (52) are not Jacobi elliptic function anymore. This corresponding to the saying of the last part that the static condition means we have degenerate branch points.

### 5.3 Reciprocal transformaton and dual viewpoint of kink-breather transition:

It is interesting to know the static kink of both case are the same, while the static breather of both case exist reciprocal transformation.

$q^{(a,b)}_{kink}(x) = 4\tan^{-1}[\sqrt{k'} \cdot sc(\frac{x}{1-k'}; k)] = 4\tan^{-1}[\frac{\sqrt{k'}}{k} \cdot sd(\frac{kx}{1-k'}; \frac{1}{k})] = 4\tan^{-1}\{\sqrt{ik_1 k_1'}\, sd[(k_1 \pm ik_1')x; k_1]\}$

$q^{(a)}_{breather}(x) = 4\tan^{-1}\{\sqrt{\frac{ik_1'}{k_1}} \cdot nc[(k_1 \pm ik_1') \cdot x; k_1]\}$

$= 4\tan^{-1}[\sqrt{k'}\, nc(\frac{kx}{1+k'}; \frac{1}{k})] = 4\tan^{-1}[\sqrt{k'} \cdot nd(\frac{x}{1+k'}; k)] = q^{(b)}_{breather}(x) \tag{60}$

where $k_1 = \frac{1}{k}$. (60) means $q^{(b)}_{breather}$ is reciprocal transformation of $q^{(a)}_{breather}$. Because we have following relations:

$$dn(u+K+iK',k) = ik' \cdot sc(u,k)$$
$$cn(u+K+iK',k) = \frac{-ik'}{k} \cdot nc(u,k)$$
$$dn(ku, \frac{1}{k}) = cn(u,k)$$
$$cn(ku, \frac{1}{k}) = dn(u,k) \tag{61}$$

We will explain the opposite sign in the denominator of (56) and (58) inside the Jacobi elliptic function, that is $\sqrt{k'} sc[\frac{x}{1-k'} + iK', k] = \frac{i\sqrt{k'}}{dn[\frac{x}{1-k'},k]}$. In addition redefine $\widetilde{k'} = -k'$ so that $\frac{i\sqrt{k'}}{dn[\frac{x}{1-k'},k]} = \frac{\sqrt{\widetilde{k'}}}{dn[\frac{x}{1+\widetilde{k'}},\widetilde{k}]}$. This means the static solutions, (56)~(58) are " the same" one but only in different range of $k$. (p.s. $k \to -k$ or $k \to \frac{1}{k}$) This unifying picture is the result of analytical continuation for $k$, too. Because

$$k \cdot sc(x,k) = sd(kx, \frac{1}{k})$$
$$cn(u+K,k) = -k' \cdot sd(u,k) \tag{62}$$

We can apply the transformation, (59), (60) to our interesting physical system in [7], [8], there we consider the system with a adiabatic parameter, so the solutions of SG can go from kink to breather and go back, ect. This is a circle circulation to get a quantum spin pump. And It is interesting to know kow and what the transformation between kink and breather when the adiabatic parameter changes. In [7], [8], we mainly discuss **case(b)** in Sec.4.We can observe that if we first do "Reciprocal transformation" to $q_{kink}^{(a)}$, like (59), make $sc$ function of kink into $sd$ function, and if we want to get breather, the $nc$ function, we want to shift a imaginary shift $iK'$. But this seem a sudden change from kink to breather, we have another "smooth" viewpoint as follows by the viewpoint of **Case(a)**.

That if we begin from $sc$ function of kink, then by (59), first we shift back $K + iK'$, so that $sc \to dn$, then we do "Reciprocal transformation" by $dn \to cn$. Finally, we add lost "$K + iK'$" to $cn$, so that $cn \to nc$, this is the inside function of $q_{breather}^{(a)}$. In this point, we may think there are no imaginary shift in this process. The mathematical reason is

$$K'(\frac{1}{k}) = k \cdot K'(k)$$
$$K(\frac{1}{k}) = k \cdot [K(k) + iK'(k)] \tag{63}$$

after we "reciprocal transformation" and add lost $K(\frac{1}{k}) + iK'(\frac{1}{k})$. The unexpected imaginary shift: $iK'$ is contained from $K(\frac{1}{k})$.

**Appendix A**: **The Floquet discriminant** $\Delta(E)$:

From [4], Fix a point $x_0$, a basis of solutions of (1) is $\{\phi_+(x,x_0,E), \phi_-(x,x_0,E)\}$, by the initial conditions at $x = x_0$:

$$\phi_+(x = x_0, x_0, E) = \begin{pmatrix} 1 \\ 0 \end{pmatrix}, \phi_-(x = x_0, x_0, E) = \begin{pmatrix} 0 \\ 1 \end{pmatrix} \tag{A.1}$$

but from the period of (2), $\phi_\pm(x+L, x_0, E)$ are aslo solutions of (1), we can expand them on the basis of $\phi_\pm(x, x_0, E)$,

$$\begin{pmatrix} \phi_+(x+L, x_0, E) \\ \phi_-(x+L, x_0, E) \end{pmatrix} = \begin{pmatrix} t_{11}(E) & t_{12}(E) \\ t_{21}(E) & t_{22}(E) \end{pmatrix} \cdot \begin{pmatrix} \phi_+(x, x_0, E) \\ \phi_-(x, x_0, E) \end{pmatrix} = T(E) \cdot \begin{pmatrix} \phi_+(x, x_0, E) \\ \phi_-(x, x_0, E) \end{pmatrix} \tag{A.2}$$

If we require the bounded behavior of $\phi_\pm(x+NL, x_0, E)$ for large N, then $|\rho_\pm(E)|=1$ where $\rho_\pm(E)$ are the eigenvalues of thr transfer matrix $T(E)$,

$$\det[T(E) - \rho(E)I] = 0 \tag{A.3}$$

Computing the determinant

$$\rho^2(E) - \Delta(E)\rho(E) + 1 = 0 \tag{A.4}$$

where $\Delta(E) \equiv$ trace of $T(E)$ is known as the Floquet discriminant.

$$\rho_-(E)\rho_+(E) = 1$$
$$\Delta(E) = \rho_-(E) + \rho_+(E) \tag{A.5}$$

$\Delta(E)$ is an imporatant quantity, we will study the relation between $\Delta(\frac{1}{16^2 E})$ and $\Delta(E)$ both in even and odd cases.

Case 1: If $u, \psi_1, \psi_2$ are even functions:

And we use the same notations as in Section 1.1. In this case, if
$\psi(x, E) = \begin{pmatrix} \psi_1(x, E) \\ \psi_2(x, E) \end{pmatrix}$ solve (4) at $E$, then $\begin{pmatrix} \psi_2(-x, \frac{1}{16^2 E}) \\ \psi_1(-x, \frac{1}{16^2 E}) \end{pmatrix}$ solve (4) at $\frac{1}{16^2 E}$. The Floquet discriminant can be represented as

$$\Delta(E) = \phi_{+,1}(L, E) + \phi_{-,2}(L, E) \tag{A.6}$$

where $\phi_\pm(x, E)$ are the basis of (1) and normalized by

$$\phi_+(x=0, E) = \begin{pmatrix} 1 \\ 0 \end{pmatrix}; \phi_-(x=0, E) = \begin{pmatrix} 0 \\ 1 \end{pmatrix} \tag{A.7}$$

we define

$$\psi_\pm(x, E) = \begin{pmatrix} e^{\frac{i}{4}u(x)}\phi_{\pm,1}(x, E) \\ e^{-\frac{i}{4}u(x)}\phi_{\pm,2}(x, E) \end{pmatrix} \tag{A.8}$$

using the initial conditions

$$\begin{pmatrix} \psi_{\pm,2}(0, \frac{1}{16^2 E}) \\ \psi_{\pm,1}(0, \frac{1}{16^2 E}) \end{pmatrix} = e^{\frac{i}{4}u(0)}\begin{pmatrix} 0 \\ 1 \end{pmatrix}$$
$$= e^{\frac{-i}{4}u(0)}\begin{pmatrix} 1 \\ 0 \end{pmatrix} \tag{A.9}$$

then the eigenfunctions are proportional

$$\begin{pmatrix} \psi_{\pm,2}(-x, \frac{1}{16^2 E}) \\ \psi_{\pm,1}(-x, \frac{1}{16^2 E}) \end{pmatrix} = e^{\pm\frac{i}{2}u(0)}\psi_\mp(x, E) \tag{A.10}$$

Then we compute

$$\Delta(\frac{1}{16^2 E}) = \phi_{+,1}(L, \frac{1}{16^2 E}) + \phi_{-,2}(L, \frac{1}{16^2 E})$$
$$= e^{-\frac{i}{4}u(L)}\psi_{+,1}(L, \frac{1}{16^2 E}) + e^{\frac{i}{4}u(L)}\psi_{-,2}(L, \frac{1}{16^2 E})$$
$$= e^{-\frac{i}{4}u(L)}e^{\frac{i}{2}u(0)}\psi_{-,2}(-L, E) + e^{\frac{i}{4}u(L)}e^{\frac{-i}{2}u(0)}\psi_{+,1}(-L, E)$$
$$= e^{\frac{i}{2}[u(0)-u(L)]}\phi_{-,2}(-L, E) + e^{\frac{i}{2}[u(L)-u(0)]}\phi_{+,1}(-L, E) \tag{A.11}$$

From $u(L) - u(0) = 2\pi M$, where $M =$" charge of $u(x)$", we find

$$\Delta(\frac{1}{16^2 E}) = (-1)^M \Delta(E) \tag{A.12}$$

Case 2: If $u, \psi_1, \psi_2$ are odd functions:

$$\begin{aligned}
\Delta(\frac{1}{16^2 E}) &= \phi_{+,1}(L, \frac{1}{16^2 E}) + \phi_{-,2}(L, \frac{1}{16^2 E}) \\
&= e^{-\frac{i}{4}u(L)}\psi_{+,1}(L, \frac{1}{16^2 E}) + e^{\frac{i}{4}u(L)}\psi_{-,2}(L, \frac{1}{16^2 E}) \\
&= e^{\frac{i}{4}u(-L)}\psi_{+,1}(-L, E) + e^{-\frac{i}{4}u(-L)}\psi_{-,2}(-L, E) \\
&= e^{-\frac{i}{2}u(L)}e^{-\frac{i}{4}u(-L)}\psi_{+,1}(-L, E) + e^{\frac{i}{2}u(L)}e^{\frac{i}{4}u(-L)}\psi_{-,2}(-L, E) \\
&= e^{-\frac{i}{2}u(L)}\phi_{+,1}(-L, E) + e^{\frac{i}{2}u(L)}\phi_{-,2}(-L, E) \\
&= (-1)^M \Delta(E)
\end{aligned} \qquad (A.13)$$

So both even and odd cases share the same relation between $\Delta(E)$ and $\Delta(\frac{1}{16^2 E})$.

**Appendix B**: **Calculate $\overleftrightarrow{\sigma}_b$ in Sec. 3.2**:

Because $\overleftrightarrow{\sigma}_c \cdot \overleftrightarrow{B}_b = \overleftrightarrow{B}_a$, then $\overleftrightarrow{\sigma}_a \cdot \overleftrightarrow{B}_a = \overleftrightarrow{\sigma}_a \cdot \overleftrightarrow{\sigma}_c \cdot \overleftrightarrow{B}_b$. We argue $\overleftrightarrow{\sigma}_b = \overleftrightarrow{\sigma}_a \cdot \overleftrightarrow{\sigma}_c$ and the multiply is $2 \times 2$ instead of $4 \times 4$.

(pf): We divide $4 \times 4$ matrices $\overleftrightarrow{\sigma}_c$ and $\overleftrightarrow{\sigma}_a$ as $\overleftrightarrow{\sigma}_c = \begin{pmatrix} a_c & b_c \\ c_c & d_c \end{pmatrix}$ and $\overleftrightarrow{\sigma}_a = \begin{pmatrix} a_a & b_a \\ c_a & d_a \end{pmatrix}$. Where each matrix elements $a_{a,c}$ to $d_{a,c}$ are $2 \times 2$ matrix. By definition,

$\overleftrightarrow{\sigma}_c \cdot \overleftrightarrow{B}_b = \frac{a_c \cdot \overleftrightarrow{B}_b + b_c}{c_c \cdot \overleftrightarrow{B}_b + d_c}$ and $\overleftrightarrow{\sigma}_a \cdot (\overleftrightarrow{\sigma}_c \cdot \overleftrightarrow{B}_b) = \frac{a_a \cdot \frac{a_c \cdot \overleftrightarrow{B}_b + b_c}{c_c \cdot \overleftrightarrow{B}_b + d_c} + b_a}{c_a \cdot \frac{a_c \cdot \overleftrightarrow{B}_b + b_c}{c_c \cdot \overleftrightarrow{B}_b + d_c} + d_a} = \frac{(a_a \cdot a_c + b_a \cdot c_c) \cdot \overleftrightarrow{B}_b + (a_a \cdot b_c + b_a \cdot d_c)}{(c_a \cdot a_c + d_a \cdot c_c) \cdot \overleftrightarrow{B}_b + (c_a \cdot b_c + d_a \cdot d_c)}$. If we define

$\overleftrightarrow{\sigma}_b = \overleftrightarrow{\sigma}_a \cdot \overleftrightarrow{\sigma}_c = \begin{pmatrix} a_a & b_a \\ c_a & d_a \end{pmatrix} \cdot \begin{pmatrix} a_c & b_c \\ c_c & d_c \end{pmatrix} = \begin{pmatrix} a_a \cdot a_c + b_a \cdot c_c & a_a \cdot b_c + b_a \cdot d_c \\ c_a \cdot a_c + d_a \cdot c_c & c_a \cdot b_c + d_a \cdot d_c \end{pmatrix}$, then

$\overleftrightarrow{\sigma}_a \cdot (\overleftrightarrow{\sigma}_c \cdot \overleftrightarrow{B}_b) = (\overleftrightarrow{\sigma}_a \cdot \overleftrightarrow{\sigma}_c) \cdot \overleftrightarrow{B}_b = \overleftrightarrow{\sigma}_b \cdot \overleftrightarrow{B}_b$. So

$$\overleftrightarrow{\sigma}_b = \overleftrightarrow{\sigma}_a \cdot \overleftrightarrow{\sigma}_c = \begin{pmatrix} \begin{pmatrix} 0 & -1 \\ 0 & 0 \end{pmatrix} & \begin{pmatrix} 2 & -1 \\ 0 & 1 \end{pmatrix} \\ \begin{pmatrix} 0 & 0 \\ -1 & -1 \end{pmatrix} & \begin{pmatrix} 1 & -1 \\ 0 & 0 \end{pmatrix} \end{pmatrix} \cdot \begin{pmatrix} \begin{pmatrix} 2 & -1 \\ 0 & 1 \end{pmatrix} & \begin{pmatrix} 0 & 0 \\ 0 & 0 \end{pmatrix} \\ \begin{pmatrix} 0 & 0 \\ 0 & 0 \end{pmatrix} & \begin{pmatrix} 1 & 0 \\ 1 & 2 \end{pmatrix} \end{pmatrix} = \begin{pmatrix} 0 & -1 & 0 & 0 \\ 0 & 0 & 1 & 0 \\ 0 & 0 & 0 & -2 \\ -2 & 0 & 0 & 0 \end{pmatrix}$$